# Detecting 3D vegetation structure with the Galileo space probe: Can a distant probe detect vegetation structure on Earth?

Christopher E. Doughty[1]*, Adam Wolf[2]

1. School of Informatics, Computing, and Cyber Systems, Northern Arizona University, Flagstaff, AZ. 86011, USA.

2. Department of Ecology and Evolutionary Biology, Princeton University, Princeton NJ 08544, USA

.

**Short title –** Detecting three-dimensional vegetation structure

# 1. Abstract

Sagan et al. (1993) used the Galileo space probe data and first principles to find evidence of life on Earth. Here we ask whether Sagan et al. (1993) could also have detected whether life on Earth had three-dimensional structure, based on the Galileo space probe data. We reanalyse the data from this probe to see if structured vegetation could have been detected in regions with abundant photosynthetic pigments through the anisotropy of reflected shortwave radiation. We compare changing brightness of the Amazon forest (a region where Sagan et al. (1993) noted a red edge in the reflectance spectrum, indicative of photosynthesis) as the planet rotates to a common model of reflectance anisotropy and found measured increase of surface reflectance of $0.019 \pm 0.003$ versus a $0.007$ predicted from only anisotropic effects. We hypothesize the difference was due to minor cloud contamination. However, the Galileo dataset had only a small change in phase angle (sun-satellite position) which reduced the observed anisotropy signal and we demonstrate that theoretically if the probe had a variable phase angle between 0-20°, there would have been a much larger predicted change in surface reflectance of $0.06$ and under such a scenario three-dimensional vegetation structure on Earth could possibly have been detected. These results suggest that anisotropic effects may be useful to help determine whether exoplanets have three-dimensional vegetation structure in the future but that further comparisons between empirical and theoretical results are first necessary.

## 2. Introduction

Carl Sagan and others published a seminal paper in *Nature* using the Galileo space probe data to see if signs of life could be detected remotely on Earth to serve as a type of "control" experiment for the new field of astrobiology [1]. Sagan et al (1993) used the Galileo data and first principles to find evidence of life on Earth based on a series of methodologies that are called by some, the "Sagan criteria for life". Since then, interest in astrobiology has grown with the advancement of these techniques [2-10]. The Kepler satellite, which has been used to discover over 962 confirmed exoplanets, has further accelerated interest in astrobiology [11, 12]. Recently, Earth-like planets as close as four light years away from Earth have been discovered within the "goldilocks zone" potentially capable of supporting life [13, 14]. However, did Sagan et al. (1993) miss anything in their original analysis that we can use as we search for other habitable planets? Here we reanalyse the original Galileo space probe data (http://pdsimg.jpl.nasa.gov/data) to test whether they missed a big potential life stage, that is, the existence of vegetation with three-dimensional structure on Earth [15]. Using the instruments of the space probe, Sagan et al (1993) found evidence on Earth for an oxygen atmosphere, an ozone layer, atmospheric methane, photosynthetic pigments, and radio signals indicative of intelligent life on Earth. However, an oxygen atmosphere and photosynthetic pigments can be signs of single-cellular life and are not sufficient evidence of abundant vegetation with three-dimensional structure. For instance, the patch of photosynthetic pigments observed by Sagan could have been "green slime," since according to isotopic evidence, single cellular photosynthetic organisms were likely present on land from 1.2 billion years ago [16]. Radio signals are clear signs of intelligence and likely signs of multicellular life. However, if we examine Earth's life history, there has been abundant multicellular life for ~500 million years, but radio signals have only emanated from Earth for about ~100 years. Therefore, we would want a technique to distinguish between multicellular and intelligent (technological) life, since the ~500 million year period since the Cambrian explosion gave rise to "endless forms most beautiful and most wonderful" [17].

How can we distinguish abundant single cellular life from abundant complex vegetation with three-dimensional structure? We hypothesize that this is possible by viewing shadows cast by objects on Earth, in the presence of weather that would erode non-living surfaces [15]. On Earth, as Sagan et al. (1993) pointed out, there is abundant water and clouds, which together are a strong sign of energetic, erosive precipitation. Gravity plus weather erodes most surfaces. In fact, >99% of geologic surfaces (i.e. orography) has an incline of < 45˚ versus ~90˚ for most trees [18]. However, despite all the natural erosion on Earth due to weather and climate, Earth has abundant shadows at low to modest solar zenith angles (that is, < 45°) because of the presence of trees. We posit that trees are the only such objects that cast abundant shadows at low to modest solar zenith angles. Trees are unique in that while they will be periodically knocked down by weather and entropy in general, as living objects they will regrow and be abundant on the planetary surface. There are periodic geologic structures on Earth like the Hoodoo formations in Utah that cast shadows at low to modest zenith angles, but we hypothesize that they will be rare on every planet with weather and climate, and never as abundant as trees in wet regions.

We propose to use shadows cast by objects on Earth at certain sun angles to determine, without any prior knowledge, whether Earth has vegetation with three-dimensional structure. Remote sensors have been quantifying shadows cast by objects on Earth for decades using a technique called the bidirectional reflectance distribution function (BRDF) [19-25]. This is the change in observed reflectance with changing view angle or illumination direction [26]. Surface albedo results if the BRDF is integrated over the entire viewing hemisphere. This technique has been extensively tested for accuracy [19, 27].

It is useful in this context to define what a shadow is, under different illumination and view geometries (Figure 1). There is the familiar *solar shadow*, that is the projection of a solid object onto the ground from the illumination of the sun, and also a *view shadow*, that is the projection of a solid object onto the ground and hidden from view from the perspective of a viewer or planetary imager. Assuming that the illumination or view are distant enough that rays are considered parallel, then as solar zenith angles ($\Omega_i$) or view zenith angles ($\Omega_v$) increase, the area of the projected shadow increases as $1/\cos(\Omega)$. When $\Omega_v = \Omega_i$ and the relative azimuth (Ø) is zero, no solar shadows will be visible to the viewer, and a pixel including objects and their background will have maximal brightness. This angle when the solar shadow and view shadow coincide is called the *hot spot*. As the solar and view angles diverge (that is phase angles Φ increase by changes in $\Omega_v - \Omega_i$ and/or Ø), then more of the solar shadow will be visible to the viewer, and pixel brightness will decrease. The most shadows will be visible within the principle plane, that is Ø = 0 or π, but in the mirror angle direction ($\Omega_v = -\Omega_i$). This angle, when the solar shadow and view shadow are most disparate, is called the *dark spot*.

In a previous paper [15], we hypothesized that under certain conditions, ΔBRF (the change in a planet's reflectance at different phase angles) could be used to determine whether exoplanets have three-dimensional vegetation structure. In this paper, we want to demonstrate proof of concept for this idea in a similar manner to how Sagan et al (1993) did so, using only the reflectance anisotropy collected from the Galileo space probe to prove that Earth has vegetation with three-dimensional structure. Here we ask whether vegetation with three-dimensional structure can be detected on Earth using data from a distant space probe (Galileo) and a common model of reflectance anisotropy [28, 29].

## 3. Materials and Methods

We use data from the solid state imaging system [30] on the Galileo probe, which took images of Earth in six wavelength bands as it passed by Earth on its way to Jupiter in December 1990. We focus on the region of the planet (the northern Amazon forest) that was demonstrated by Sagan to have a red edge in the reflectance spectrum indicative of photosynthetic pigments. Do the photosynthetic pigments seen by Sagan et al. (1993) have three dimensional structure, or are they structurally more similar to "green slime"? We view the reflectance for this region in the near infrared (NIR) images as the region moves from a solar zenith angle of 16.6° to one close to zero (sun overhead) (Figure 2A to B) for cloud free (blue reflectance <0.1) regions with abundant vegetation (with high Normalized Difference Vegetation Index - NDVI >0.5). We sample NIR reflectance (band 5 – 756 nm) for a square region of 81 pixels (nominal pixel scale is 114 km$^2$ at sub-spacecraft point) and calculate the mean and standard error for the difference between the pixels. The Galileo probe only has geometry data for the center pixel of each scene (marked with an "X" in Figure 1a with the geometry data in Table 1). Based on this center pixel geometry, we calculate the geometry for the patch of Amazon forest that Sagan et al (1993) saw evidence of photosynthetic pigments for the two images shown (red boxes in Figure 2a to b). We show the geometry for these two areas in polar coordinates in Figure 3b which shows how $\Omega_i$ and $\Omega_v$ can change with Φ remaining constant.

We use the following equation to predict reflectance at different sun angles [28, 29], which accounts for the shading effects described above:

$$R(\Omega_i, \Omega_v, \emptyset) = k0 + k1 F1(\Omega_i, \Omega_v, \emptyset) + k2 F2(\Omega_i, \Omega_v, \emptyset) \quad \text{Equation 1}$$

This equation calculates reflectance (R) for a given territory, as seen from a viewer at zenith angle $\Omega_v$, and illuminated by the sun at zenith angle $\Omega_i$, where the relative azimuth is given by Ø. The model expresses reflectance as the sum of two "kernels", that is 3-D functions, namely a geometric kernel (F1), which models the shadows cast by randomly distributed spheroids above a flat Lambertian surface [19]; and a volumetric kernel (F2), which models a theoretical turbid vegetation canopy with high leaf density [29]. The coefficients $k_0$, $k_1$, $k_2$ are biome dependant constants inverted from multi-angle reflectance data collected in the global POLDER satellite database [28]. In our circumstances, we would not know the vegetation type a priori, so we use three vegetation types, the average values for terrestrial vegetation in the NIR (765 nm in POLDER) of $k_0 = 0.264$, $k_1 = 0.027$, and $k_2 = 0.363$, then broadleaf evergreen forest ($k_0 = 0.3257$, $k_1 = 0.0481$, and $k_2 = 0.6412$), and no vegetation (snow/ice but with a $k_0$ of desert - $k_0 = 0.296$, $k_1 = 0.0476$, and $k_2 = 0.2403$) [28]. Note that in computing $\Delta$BRF, the constant term $k_0$ drops out, such that the overall difference in brightness between a desert and forest becomes irrelevant, leaving only differences in 3D structure. In this sense, the reflectance anisotropy of the no vegetation becomes a proxy for that of "green slime".

We then calculate the slope of the increased NIR reflectance as the planet turns ($\Delta$BRF following the methodology of Wolf et al 2010 and applied previously to detecting vegetation structure in Doughty and Wolf 2010). BRF refers to the Bidirectional Reflectance Factor, a measure of reflectance anisotropy that is the ratio of radiation exiting a solid angle of a given surface divided by that of a perfect Lambertian surface. $\Delta$BRF is then the difference of the BRF for two different views of the same pixel at different Φ, and has been demonstrated to be sensitive to parameters that govern geometric optics, including the crown radius, tree number density and ground cover (Wolf et al. 2010).

## 4. Results

The Galileo SSI data shown in the red boxes of Figure 2a and b initially had a solar zenith angle ($\Omega_i$) of 16.6° (Figure 1a) which changed to 0° (Figure 1b) approximately 11 scenes later (Galileo SSI data is collected every 6 minutes and every scene shows the Earth rotated 1.5° so 16.6° = 11 scenes later). At a solar zenith angle of 0° the sun would be directly overhead and solar shadows would be minimized [20, 21, 23], but because the probe is not in line with the sun (the phase angle >0), shadows would still be visible when the forest is viewed from the side and the total change in surface reflectance would not be expected to be large. Over the 12 scenes, NIR reflectance increased by an average of 0.0017±0.0003 per scene (Figure 3c) and over 11 scenes (between Figure 1a and b), NIR reflectance increased by 0.019 ± 0.003.

Does this change in reflectance match what we would expect based on directional anisotropic effects from a changing solar and view zenith angle? Based on the parameters of the reflectance anisotropy model and the geometry listed in Table 1, we calculate the reflectance anisotropy in the NIR for the red box in Figure 2 A and B (Figure 3a and b). We predict that surface reflectance for a theoretical broadleaf evergreen forest would increase by 0.0072 (green line in Figure 3c) between these two scenes due to anisotropic effects. This fairly small predicted increase in surface reflectance is because there is no change in phase angle between the two scenes (Table 1). With no vegetation (brown line Figure 3c), we estimate a change in surface reflectance of 0.0018 between the two scenes, a 4-fold decrease between the more structured scene (broadleaf) versus the least structured (no vegetation).

We predicted a small increase in surface reflectance of 0.0072 (green line in Figure 3c) but measured an increased surface reflectance of 0.019±0.003 with the Galileo data. We cannot be sure what caused the difference between the predicted and measured surface reflectance but we hypothesize that it could be a minor increase in cloud cover which led to the increased albedo. Since the nominal pixel scale is 114 km$^2$, even a small increase in cloud percentage could account for the observed difference. This issue could be resolved by watching the same scene over multiple days, but this is unfortunately not an option for the Galileo data as only one day of data is available. A small change in cloud cover could swamp the anisotropic signal because with no change in phase angle the predicted signal was very small.

If the probe left with a varying phase angle of 0-20° instead of a continuous phase angle of 35.2°, like it did, how would surface reflectance theoretically vary? We show an example of such a flight path (Figure 4a) and its impact on theoretical BRF for just the geometric kernel (F1) (Figure 4b) between forested and no vegetation regions. Our results suggest that over such a varying phase angle, there would be a change in surface reflectance of up to 0.06 just from geometric effects. With such a large predicted signal, minor cloud contamination would be less of an issue.

## 5. Discussion

Could data from Galileo be used to unambiguously detect vegetation with three-dimensional structure on Earth? The predicted increase in surface reflectance due to anisotropic effects did not match the measured increase from the Galileo data likely due to minor cloud contamination issues. We therefore conclude that with this particular dataset, vegetation structure could not have been unambiguously detected on Earth.

However, we still posit that the method has promise for detecting three dimensional vegetation structure on exoplanets (with proper future technology). The problem with the Galileo data is that the Galileo flight path was not planned with the goal of identifying vegetation structure and therefore had a very small phase angle. However, we demonstrate that theoretically if the probe left with a varying phase angle of 0-20˚ instead of a continuous phase angle of 35.2˚, the change in surface reflectance would have been up to 0.06 which likely would have been sufficient to have identified vegetation structure on Earth.

A previous study found that 40% of variance in ΔBRF was explained by geometric components such as tree number, crown size, and ground cover (Wolf et al. 2010). These geometric components would be key in trying to distinguish a slime covered region of planet versus a forest covered one. Such large anisotropic effects are not unusual and they have clearly been shown by POLDER [20] and more recently by MODIS [31]. For instance, the Amazon forest, the region shown by Sagan et al. (1993) to have photosynthetic pigments, was demonstrated to have strong directional anisotropic effects [31] with changing seasonal viewing angles increasing NIR reflectance by ~0.05.

Could purely physical phenomena mimic a peaked pattern of brightness similar to that of BRDF? Increased cloud cover could mimic the increased brightness, but at the resolution of the Galileo probe we can broadly screen out large clouds (although there are undoubtedly small or thin clouds in the region). Repeated viewing cycles could also likely screen out these smaller clouds. As we mentioned previously, no geologic features, in regions with active weather could likely replicate the BRDF effect on that scale. Ice can have periodic steepness like forests, but this steepness will be much rarer and we can also screen out ice by its unique reflectance spectrum.

In a previous paper [15], we used the same semi-empirical BRDF model to simulate Earth and a hypothetical tree-less planet with liquid water as viewed as a single pixel from a great distance at different planetary phase angles with both simulated and real cloud cover. Even if the entire planetary albedo were viewed as a single pixel, the rate of increase of albedo as a planet approaches full illumination would be comparatively greater on Earth than the hypothetical tree-less planet with liquid water. We theorized that anisotropic effects could theoretically detect tree-like multi-cellular life on exoplanets in up to 50 of the closest stellar systems under the right conditions. However, our comparisons with empirical data suggest that anisotropic effects can be difficult to detect and a large variation in phase angle may be necessary for the technique to work.

Why is this important? Stepping away from our thought experiment, it is obvious that Earth has three-dimensional vegetation structure, and also obvious that these forests cast shadows in a manner explained by BRDF theory, but the brilliance of the Sagan et al (1993) paper was using the Galileo data as a test case for remote detection of life on exoplanets. Now that we have discovered a close (4 light years away) exoplanet that is may contain liquid water, we want to know whether it has climate suitable for life, then whether it has simple life, then complex life, then intelligent life. Sagan et al (1993) gave the roadmap for three of these four characteristics and with this paper and previous work [15], we hope to now add a fourth.

**Table 1** – Geometry for the center point (x) in figure 1a (given by Galileo) and estimated geometries for the red boxes in figure 1a (Amazon A) and figure 1b (Amazon B).

|  | Latitude | Longitude | Emission angle (view zenith angle) | Incidence angle (solar zenith angle) | Phase angle | Relative azimuth angle |
|---|---|---|---|---|---|---|
| Center pixel (x) | -23 | 70 | 18.6 | 16.6 | 35.2 | 182 |
| Amazon A | 0 | 70 | 18.6 | 16.6 | 35.2 | 145 |
| Amazon B | 0 | 70 | 35.2 | 0 | 35.2 | 164 |

## 6. Figures

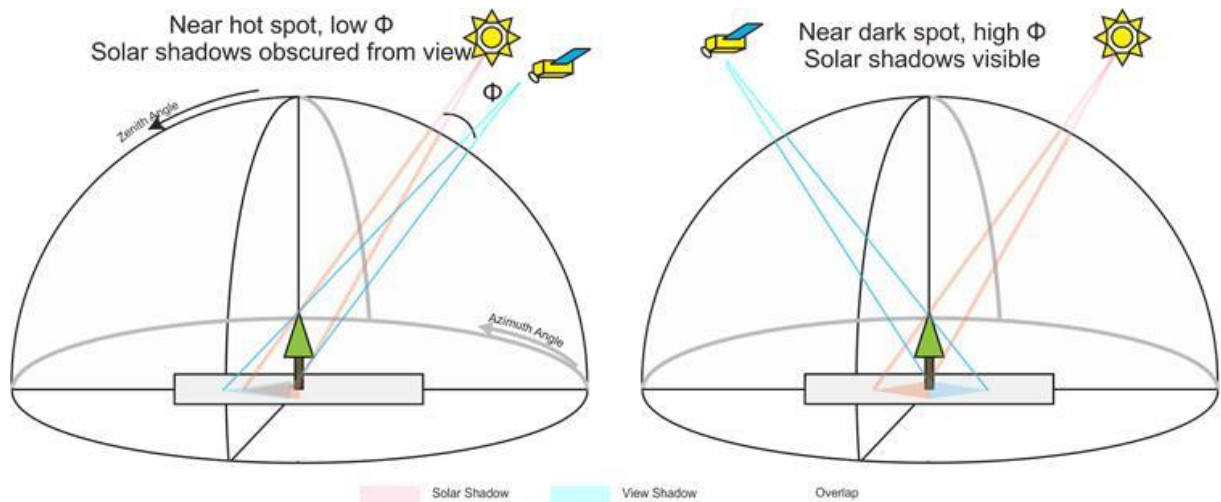

**Figure 1** – Perspective view diagram showing the geometry for a tree viewed at close to the hot spot (left) and close to the dark spot (right). Φ indicates phase angle.

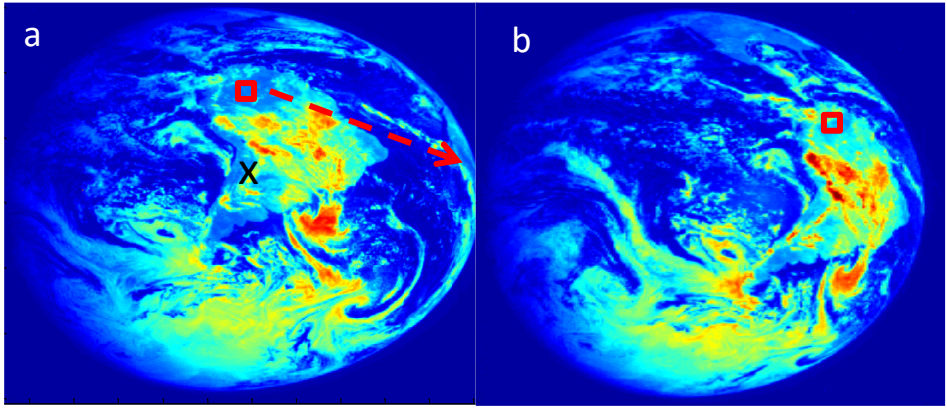

**Figure 2** – Earth as viewed from the Galileo probe at (a) 3:12 GMT December 11, 1990 and again at (b) 4:25, 12 scenes later. The red arrow shows the direction of rotation. The red square is a 9x9 pixel (~10,000 km$^2$) patch of cloud free Amazon forest.

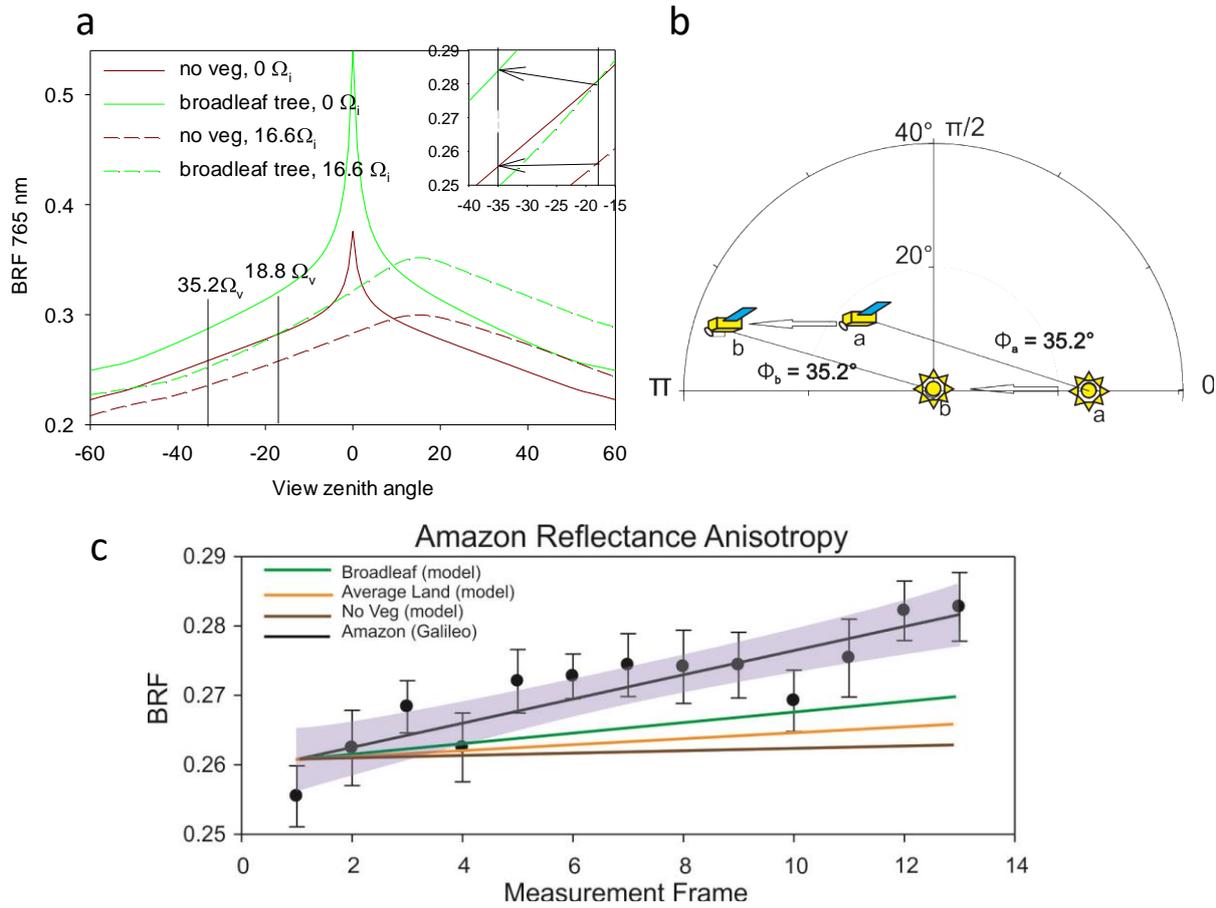

**Figure 3 -** (a) Change in estimated surface reflectance between Figure 1a and b using a BRDF model and the geometries listed in Table 1 for desert (brown) and broadleaf tree (green). The black arrows in the inset in the top corner shows the predicted change in surface reflectance for broadleaf tree and desert (b) The position of the satellites in relation to the sun for positions A and B is shown in polar coordinates. Viewing geometry with the azimuth expressed in the outer circle (0-pi) and the zenith angle expressed as its projection onto the inner circle. (c) Mean NIR reflectance for the Amazon patch for 12 images between A and B (error bars are standard deviation, and grey area is the prediction band) with estimated slope (calculated in d) for desert (brown), average land (yellow), broadleaf tree (green), and using the Galileo data (black).

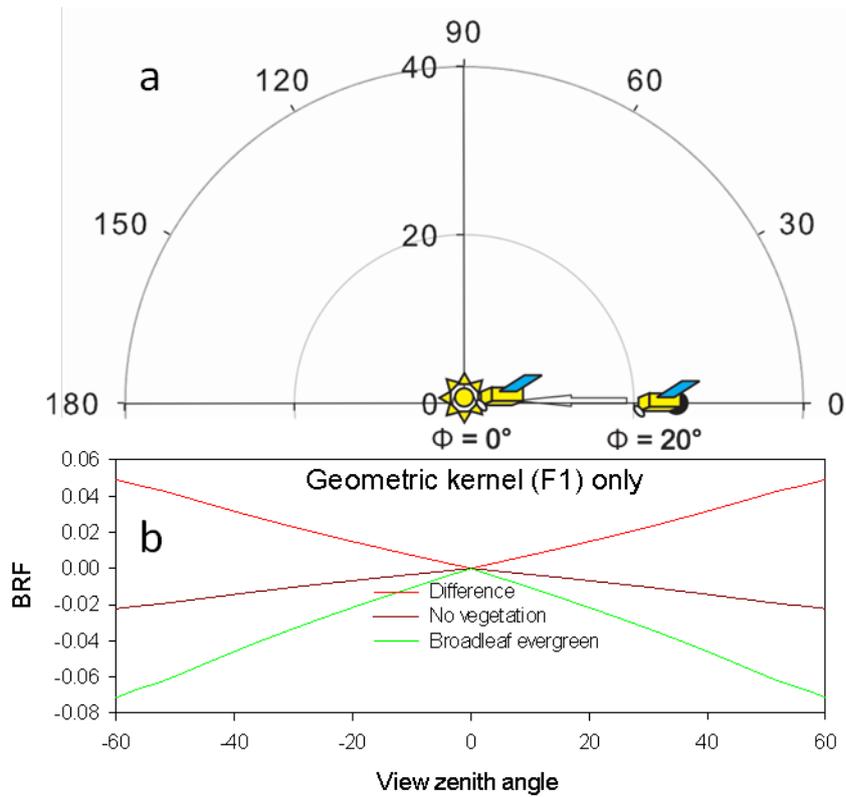

**Figure 4** – (a) Diagram showing the best potential flight path for Galileo to maximize the chance of detecting three-dimensional vegetation structure moving in the principle plane from a Φ of 20 to 0˚. (b) Estimate of changing BRF for the geometric kernel (F1) of our BRDF model for broadleaf evergreen (green), no vegetation (brown), and the difference (red). This kernel (F1) is added to the other kernel (F2) plus $K_0$ to estimate BRF.